\documentclass[aps,twocolumn,pra,showpacs,floatfix]{revtex4}

\usepackage{dcolumn}

\begin{document}

\title{\bf Atomic electric dipole moments of He and Yb induced by nuclear Schiff moments}
\author{V.A. Dzuba}
\affiliation{School of Physics, University of New South Wales,
Sydney 2052, Australia}
\author{V.V. Flambaum}
\affiliation{School of Physics, University of New South Wales,
Sydney 2052, Australia}
\author{J.S.M. Ginges}
\affiliation{School of Physics, University of New South Wales,
Sydney 2052, Australia}

\date{\today}

\begin{abstract}

We have calculated the atomic electric dipole moments (EDMs) $d$ of $^3$He and $^{171}$Yb induced by 
their respective nuclear Schiff moments $S$. Our results are 
$d(^3{\rm He})= 8.3\times 10^{-5}$ and $d(^{171}{\rm Yb})= -1.9$ in units 
$10^{-17}S/(e\,{\rm fm}^3)\,e\,{\rm cm}$.  
By considering the nuclear Schiff moments induced by the parity and time-reversal violating nucleon-nucleon 
interaction we find 
$d(^{171}{\rm Yb})\sim 0.6d(^{199}{\rm Hg})$. For $^{3}$He the nuclear EDM
coupled with the hyperfine interaction gives a larger atomic EDM than the Schiff moment. 
The result for $^3$He is required for a neutron EDM experiment that is under development, where $^3$He  
is used as a comagnetometer. We find that the EDM for $^3$He is orders of magnitude smaller than the neutron EDM. 
The result for $^{171}$Yb is needed for the planning and interpretation of experiments 
that have been proposed to measure the EDM of this atom.

\end{abstract}


\maketitle

\section{Introduction}

There are a number of experiments underway to measure the time-reversal violating electric dipole moment 
(EDM) of the neutron and of various atoms and molecules.
The measurement of a non-zero EDM would signal the 
presence of new sources of CP violation beyond the standard model, standard model EDMs being undetectably small 
\cite{PospelovRitz_review}.

The best limit on the neutron EDM is $|d_n|<2.9\times 10^{-26}\,e\,{\rm cm}$ (90\% c.l.) \cite{ILLneutronEDM} 
and that on an atomic EDM has been obtained for Hg, $|d(^{199}{\rm Hg})|<2.1\times 10^{-28}\,e\,{\rm cm}$ (95\% c.l.) 
\cite{mercuryEDM}. 
The result for Hg constrains new physics scenarios largely in the nuclear sector, since Hg is a diamagnetic atom 
(total electronic angular momentum $J=0$) and in lowest order the electric field couples to the nuclear spin. 
The Hg EDM is induced most efficiently from 
a nuclear Schiff moment; this is essentially a residual nuclear EDM (largely screened by atomic electrons) 
and is non-zero due to the finite size of the nucleus \cite{Schiff}.    
 
In this work we calculate the atomic EDMs for $^3$He and $^{171}$Yb induced by their respective nuclear Schiff moments. 
The result for $^3$He is required for a neutron EDM experiment that is under development, where $^3$He is used 
as a comagnetometer \cite{Ito,GolubLam}. 
There are proposals to measure an atomic EDM in Yb \cite{takahashi,fortson,natarajan} and preliminary work 
towards such a measurement is underway \cite{ybkyoto}.  
Determination of the size of the EDM induced by the Schiff moment is needed to gauge the relative sensitivity of 
the experiment and will be required for interpretation of the measurement. We note that a calculation for the 
Yb EDM induced by the tensor parity and time-reversal violating electron-nucleon interaction was performed in 
Ref. \cite{das}, however this is not a leading contribution in popular models of CP-violation \cite{PospelovRitz_review}.

\section{Method of calculation}

The method we use in the current work is the same as one of the methods we used 
in our earlier work \cite{DFGK} to calculate the EDMs of Hg, Xe, Rn, and Ra induced by 
nuclear Schiff moments. 
In the first method used in that work, which is the one adopted here, 
the atoms are considered as closed-shell atoms and the EDMs are
calculated using the relativistic Hartree-Fock (HF) and 
time-dependent Hartree-Fock (TDHF) methods. All atomic electrons 
contribute to the self-consistent potential. This approach
is called the $V^{N}$ approximation, where $N$ is the total number of electrons. 
For the atoms Hg and Ra, another method was also used, where HF and TDHF calculations 
were performed for atomic cores consisting of $N-2$ electrons, with the two valence electrons removed. 
The valence electrons were included at the next stage of the calculations where the  
configuration interaction, supplemented by many-body perturbation theory for valence electrons 
with the core, was implemented 
using the approach set out in \cite{DFK}.
It was demonstrated in our work \cite{DFGK} that the two methods give very close results for the 
respective EDMs of Hg and Ra. 
Therefore, in the present work we use only the simpler $V^N$ approximation. 
We briefly outline the method of calculation below.

The atomic EDM induced in the many-body state $N$ by the parity and time-reversal violating 
($P,T$-odd) interaction $H_{PT}$ can be expressed as
\begin{equation}
\label{eq:fulledm}
d=2\sum _{M}\frac{\langle N|H_{PT}|M\rangle
\langle M|D_{z}|N\rangle}{E_{N}-E_{M}} \ ,
\end{equation}
where the sum $M$ runs over a complete set of many-body states,
$E_{N}$ and $E_{M}$ are atomic energies, and
$D_{z}$ is the atomic electric dipole operator.
The $P,T$-odd interaction Hamiltonian $H_{PT}$ has the form
\begin{equation}
H_{PT}=\sum _{i}h_{PT}^{i}=-e\sum _{i}\varphi ({\bf R}_{i}),
\end{equation}
where $\varphi$ is the electrostatic potential produced by the
nuclear Schiff moment $S$ which mixes states of opposite parity. 
The form for this potential that is suitable for relativistic atomic calculations 
is \cite{FG02}
\begin{equation}
\label{eq:phi}
\varphi ({\bf R})=-\frac{3 {\bf S}\cdot {\bf R}}{B}\rho (R)\ ,
\end{equation}
where $B=\int \rho (R)R^{4}dR$ and $\rho (R)$ is the nuclear density.

In the $V^{N}$ approximation, we can write the atomic EDM induced
by the Schiff moment as
\begin{equation}
\label{eq:edmn}
d=2\sum _{n}\langle \delta n_{PT}|d_{z}|n\rangle \ ,
\end{equation}
where the sum runs over the relativistic Hartree-Fock core states $|n\rangle$,
$d_{z}$ is the single-particle dipole operator,
and $|\delta n _{PT}\rangle$ denotes
the correction to the state $|n\rangle$ due to the $P,T$-odd
Hamiltonian $h_{PT}$.
The correction $|\delta n _{PT}\rangle$ can be expressed as
\begin{equation}
\label{eq:deltansum}
|\delta n _{PT}\rangle =
\sum _{\alpha}\frac{\langle \alpha |h_{PT}|n\rangle}
{\epsilon _{n}-\epsilon _{\alpha}}|\alpha\rangle \ ,
\end{equation}
where $|\alpha\rangle$ corresponds to an excited state.
It is found by solving the equation
\begin{equation}
\label{eq:deltan}
(h_{0}-\epsilon _{n})|\delta n_{PT}\rangle
= -h_{PT}|n\rangle \ .
\end{equation}
(Equivalently, one may calculate the correction to $|n\rangle$ from the
electric dipole (E1) field and take the matrix element of the weak Hamiltonian to obtain $d$.)

In the $V^N$ approximation, polarization of the core due to the fields $h_{PT}$ and $d_{z}$ 
is accounted for by including the polarization due to one field using the TDHF method, 
e.g. by replacing $h_{PT}$ in (\ref{eq:deltan}) by $h_{\tilde{PT}}=h_{PT}+\delta V_{PT}$, since 
$\sum_{n}\langle \delta n_{\tilde{PT}}|d_{z}|n\rangle
=\sum _{n}\langle \delta n_{PT}|
d_{z}+\delta V_{d}|n\rangle$.

As a test of our wave functions we have performed calculations for the
ionization potentials and the scalar polarizabilities $\alpha$ of the ground-state for each atom 
and compared them with the available experimental data.
The polarizability has the same form as the EDM, Eq.\ref{eq:fulledm},
\begin{equation}
\alpha =-2\sum _{M}
\frac{|\langle N|D_{z}|M\rangle |^{2}}{E_{N}-E_{M}} \ ,
\end{equation}
with the operator $H_{PT}$ in
(\ref{eq:fulledm}) replaced by the dipole operator $D_{z}$.
So the scalar polarizabilities are calculated simply by replacing the
correction $|\delta n_{PT}\rangle$ due to the $P,T$-odd field by
the correction $|\delta n_{d}\rangle$ due the E1 field.

\section{Results of atomic calculations}

In Table \ref{tab:ip} we present our results for the ionization potentials and scalar polarizabilities 
for He and Yb alongside the available experimental data. It is seen that the results for helium are in 
good agreement with experiment, while for Yb a deviation of about 15\% is seen for the ionization 
potential. 

\begin{table}[h]
\caption{Ionization potentials (IP, in ${\rm cm}^{-1}$)
and scalar polarizabilities ($\alpha$, in $a_B^3$) for He and Yb}
\label{tab:ip}
  \begin{ruledtabular}
\begin{tabular}{lrrrrrr}
  & \multicolumn{2}{c}{IP} & & \multicolumn{3}{c}{$\alpha$} \\
         & HF & Exp. & & HF & TDHF & Exp. \\
\hline
 He & 201472 & 198311\tablenotemark[1] && 0.997 & 1.32 & 1.383\tablenotemark[2] \\
 Yb & 43130 & 50443\tablenotemark[3] && 124   & 179  &  \\
\end{tabular}
  \end{ruledtabular}
\tablenotetext[1]{Ref. \cite{martin}}
\tablenotetext[2]{Ref. \cite{radtsig}}
\tablenotetext[3]{Ref. \cite{ybIP}}
\end{table}

Results for the atomic EDMs for He and Yb induced by their respective nuclear Schiff moments are 
listed in Table \ref{tab:dvn}. For comparison and easy reference we have also presented in the table 
the results from our previous work \cite{DFGK}, calculated in the approximation $V^N$. 
The He EDM is very small, $\sim 10^{-21}(S/(e~{\rm fm}^{3}))~e~{\rm cm}$, and is stable with 
inclusion of core polarization, the value changing by $\sim 10\%$. 

As with Hg and Ra \cite{DFGK}, the largest contribution to the atomic EDM for Yb 
comes from the outer $s$ electrons and this differs in sign to the overall contribution from the core. 
The final result for Yb is about 60\% the size of the Hg EDM (in terms of their Schiff moments). 
Our numerical result for Yb agrees with an estimate obtained by scaling the Hg result, 
$d(^{171}{\rm Yb})\sim d(^{199}{\rm Hg})(SZ^2R)_{\rm Yb}/(SZ^2R)_{\rm Hg}\sim -1.6 S({\rm Yb})/S({\rm Hg})$, 
in the units of Table \ref{tab:dvn}, $Z$ is the nuclear charge, and 
$R$ is a relativistic enhancement factor; see Refs. \cite{SFK1984} and 
\cite{SAF,FG02}, respectively, where the parametric dependence was found and more recently applied.

One may be concerned about the huge corrections to Yb coming from inclusion of core polarization, the final 
value being over four times the HF value.
We remind the reader that in our previous work \cite{DFGK} we also 
saw such corrections (for Ra and to a lesser extent Hg, see Table \ref{tab:dvn}). 
In Ref. \cite{DFGK} an entirely different method for the calculation of many-body corrections 
was also carried out and yielded results for the EDMs of Hg and Ra that differ from those obtained 
in the $V^N$ approximation by less than 10\%. 

We expect that the TDHF value for the EDM of Yb is accurate to about 20-30\%, the result for 
He more accurate.
 
\begin{table}[h]
\caption{Atomic EDMs $d$ induced by respective nuclear Schiff moments $S$ 
in the HF and TDHF approximations, units are $10^{-17}(S/(e~{\rm fm}^{3}))~e~{\rm cm}$. 
Results for He and Yb are from the current work; others are from our 
previous work Ref. \cite{DFGK}.} 
\label{tab:dvn}
  \begin{ruledtabular}
\begin{tabular}{rlcc}
Z & Atom & HF & TDHF \\
\hline
 2 & He & $0.743 \times 10^{-4}$  & $0.826 \times 10^{-4}$ \\
54 & Xe &  0.289 & 0.378 \\
70 & Yb & -0.416 & -1.91 \\
80 & Hg & -1.19 & -2.97 \\
86 & Rn &  2.47 &  3.33 \\
88 & Ra & -1.85 & -8.23 \\
\end{tabular}
  \end{ruledtabular}
\end{table}

\section{Atomic EDMs in terms of the $P,T$-odd nucleon-nucleon interaction}

We'd like to express the atomic EDMs for He and Yb in terms of a more fundamental parameter, 
in particular, the parameter specifying the strength of the $P,T$-violating nucleon-nucleon interaction $\eta_{NN}$;
we are interested in this interaction because it leads to the largest Schiff moments. 
This will give a better idea of the relative sensitivities of various atomic EDMs to fundamental physics. 
For instance, an order of magnitude enhancement of the nuclear Schiff moment may occur for deformed nuclei with 
close levels of opposite parity \cite{SFK1984} (see also Refs. \cite{Feinberg,HH} where nuclear EDM enhancement was 
considered); even more spectacular is the orders of magnitude enhancement 
that may arise in nuclei with octupole deformation \cite{AFS,EFH}. 

The ground state of $^{171}_{70}$Yb has quantum numbers $J^{\pi}=1/2^{-}$. While this nucleus is deformed, there 
are no opposite parity levels, $1/2^{+}$, close to the ground state and therefore there is no close-level 
enhancement. Nuclei of $^{171}$Yb and $^{199}$Hg have the same quantum numbers and similar magnetic moments. 
In the shell model we consider the ground states to have an unpaired neutron in state $p_{/12}$.  
Therefore we may expect
\begin{equation}
 S(^{171}{\rm Yb})\approx S(^{199}{\rm Hg})=-1.4\times 10^{-8}\eta_{np}\, e\,{\rm fm}^3 \nonumber
\end{equation}
so that the atomic EDM is
\begin{equation}
d(^{171}{\rm Yb})\approx 0.6 d(^{199}{\rm Hg}) \approx 3 \times 10^{-25}\eta_{np}\,e\,{\rm cm} \ .
\end{equation}

The nuclear ground state for $^3$He is $J^{\pi}=1/2^+$ and the magnetic moment is very close ($\sim 10\%$) 
to that of the neutron; we consider that in the ground state there is an unpaired neutron 
in the state $s_{1/2}$. 
According to Ref. \cite{SFK1984}, the nuclear Schiff moment for $^3$He is 
$S(^3{\rm He})\sim 0.1\times 10^{-8}\eta_{np}\,e\,{\rm cm}$. The Schiff moment scales with atomic mass $A$ as 
$A^{2/3}$. Scaling from $^{199}$Hg, $S(^3{\rm He})\sim S(^{199}{\rm Hg})\times (3/199)^{2/3}\sim 0.1\times 10^{-8}\eta_{np}
\,e\,{\rm cm}$, so the result of Ref. \cite{SFK1984} looks reasonable. The induced atomic EDM will then be 
$d(^3{\rm He})\approx 1\times 10^{-30}\eta_{np}\,e\,{\rm cm}$. 

However, for very light atoms the finite size effect (Schiff moment) does not lead to the largest atomic EDMs. 
Another way to violate the screening of the nuclear EDM is to take account of magnetic fields. 
Considering the hyperfine interaction, an order of magnitude estimate for the atomic EDM arising from a nuclear 
EDM is \cite{Khriplovich} 
\begin{equation}
d_{\rm atom}(^3{\rm He})\sim Z\alpha^2\frac{m_e}{m_p}d_{\rm nuc}(^3{\rm He}) \ , \nonumber
\end{equation}
where $Z$ is the nuclear charge, $\alpha$ is the fine structure constant, $m_e$ and $m_p$ are electron and 
proton masses, and the nuclear EDM is denoted $d_{\rm nuc}(^3{\rm He})$. 
A more involved estimate was performed earlier by Schiff who obtained 
$|d(^3{\rm He})| = 1.5\times 10^{-7}d_{\rm nuc}(^3{\rm He})$ \cite{Schiff}. 
Using the value 
$d_{\rm nuc}(^3{\rm He}) \sim 1\times 10^{-21}\eta_{np}\,e\,{\rm cm}$ from Ref. \cite{SFK1984}, the size 
of the induced atomic EDM is then
\begin{equation}
|d(^3{\rm He})|\sim 1.5\times 10^{-28}\eta_{np}\,e\,{\rm cm} , 
\end{equation}
larger than that induced by the nuclear Schiff moment.  
 
The size of the neutron EDM induced by the nucleon-nucleon interaction is  
$d_n=0.5\times 10^{-23}\eta_{np}\,e\,{\rm cm}$ \cite{SFK1984}, and therefore it is seen that the atomic 
EDM of $^{3}$He is negligibly small compared to the neutron EDM $d_n$, 
\begin{equation}
d(^3{\rm He})\sim 3\times 10^{-5}d_n \ .
\end{equation}

\section{Summary}

We have calculated the atomic EDMs for $^3$He and $^{171}$Yb induced by their respective nuclear Schiff 
moments with the results $d(^3{\rm He})= 8.3\times 10^{-5}$ and $d(^{171}{\rm Yb})= -1.9$ in units 
$10^{-17}(S/(e\,{\rm fm}^3))\,e\,{\rm cm}$. The accuracy is about 20-30\% for Yb and is better for He.
We also estimated the sizes of the nuclear Schiff moments induced by the $P,T$-violating nucleon-nucleon 
interaction. We find $d(^{171}{\rm Yb})\sim 0.6d(^{199}{\rm Hg})$. For $^{3}$He the nuclear EDM
coupled with the hyperfine interaction gives a larger atomic EDM than the Schiff moment. 
Nevertheless, the helium EDM is orders of magnitude smaller than the neutron EDM and therefore may be 
neglected in the neutron experiment \cite{Ito,GolubLam}.

\section*{Acknowledgments}

We thank B. Filippone for motivating our work for $^3$He.
This work was supported by the Australian Research Council.

\end{document}